\documentclass[conference]{IEEEtran}
\IEEEoverridecommandlockouts

\newcommand{\paperTitle}{Challenges and Opportunities in Rapid Epidemic Information Propagation with Live Knowledge Aggregation from Social Media}
\newcommand{\paperKeywords}{TODO}
\newcommand{\paperAuthors}{Calton Pu, Abhijit Suprem, Rodrigo A. Lima}
\usepackage[hyphens]{url}
\usepackage{xstring}

\usepackage[usenames,dvipsnames]{xcolor}
\definecolor{linkcolor}{HTML}{647382}
\definecolor{citecolor}{HTML}{647382} %
\definecolor{urlcolor}{rgb}{0.4,0.2,0.2}
\definecolor{sqlcolor}{HTML}{965d67}
\definecolor{smtcolor}{HTML}{5d968c}

\usepackage[breaklinks]{hyperref}
\hypersetup{%
	pdfauthor = {\paperAuthors},
	pdftitle = {\paperTitle},
	pdfkeywords = {\paperKeywords},
	bookmarksopen = {true},
	colorlinks=true,
	citecolor={urlcolor},
	linkcolor={linkcolor},
	urlcolor={citecolor},
	pdfborder={ 0 0 0 }
}

\usepackage[usenames,dvipsnames]{xcolor}
\usepackage{amsmath,amsopn,amssymb}
\usepackage{listings}
\usepackage[caption=false]{subfig}
\usepackage{endnotes,microtype,xspace,graphicx,fancyvrb,multirow}
\usepackage{supertabular,booktabs}
\usepackage{array,underscore,relsize}
\usepackage[T1]{fontenc}
\usepackage{times}
\usepackage{fancyhdr,lastpage}
\usepackage{enumitem}
\usepackage{balance}
\usepackage{booktabs}
\usepackage{pifont}
\usepackage{listings}
\usepackage{multirow}
\usepackage[scaled]{beramono}
\usepackage{tabularx}
\usepackage{wrapfig}
\usepackage{lipsum}

\usepackage{stmaryrd}
\usepackage{ltablex}
\usepackage{mathtools}
\usepackage{mathrsfs}
\usepackage{ dsfont }

\usepackage{amsmath}
\usepackage{algpseudocode}
\usepackage[ruled,linesnumbered, vlined]{algorithm2e}
\usepackage{enumitem}
\usepackage{algpseudocode}

\usepackage{multirow}
\usepackage[normalem]{ulem}
\useunder{\uline}{\ul}{}

\usepackage[capitalize,noabbrev]{cleveref}

\def\BibTeX{{\rm B\kern-.05em{\sc i\kern-.025em b}\kern-.08em
    T\kern-.1667em\lower.7ex\hbox{E}\kern-.125emX}}

\newcommand{\squishitemize}{
 \begin{list}{$\bullet$}
  { \setlength{\itemsep}{0pt}
     \setlength{\parsep}{3pt}
     \setlength{\topsep}{3pt}
     \setlength{\partopsep}{0pt}
     \setlength{\leftmargin}{1.95em}
     \setlength{\labelwidth}{1.5em}
     \setlength{\labelsep}{0.5em} } }

\newcounter{Lcount}
\newcommand{\squishlist}{
    \begin{list}{\arabic{Lcount}. }
   { \usecounter{Lcount}
        \setlength{\itemsep}{0pt}
        \setlength{\parsep}{3pt}
        \setlength{\topsep}{3pt}
        \setlength{\partopsep}{0pt}
        \setlength{\leftmargin}{2em}
        \setlength{\labelwidth}{1.5em}
        \setlength{\labelsep}{0.5em} } }

\newcommand{\squishend}{\end{list}}

\SetCommentSty{mycommfont}

\newcommand{\thickhline}{%
	\noalign {\ifnum 0=`}\fi \hline height 1pt
	\futurelet \reserved@a \@xhline
}

\makeatletter
\newcommand{\linebreakand}{%
\end{@IEEEauthorhalign}
\hfill\mbox{}\par
\mbox{}\hfill\begin{@IEEEauthorhalign}
}
\makeatother

\newcommand{\edna}{EDNA\xspace}
\newcommand{\ednajob}{EDNA Job\xspace}

\begin{document}

\title{\paperTitle}

\author{

	\IEEEauthorblockN{Calton Pu}
	\IEEEauthorblockA{\textit{School of Computer Science} \\
	\textit{Georgia Institute of Technology}\\
	Atlanta, USA\\
	calton.pu@cc.gatech.edu}

	\and

\IEEEauthorblockN{Abhijit Suprem}
\IEEEauthorblockA{\textit{School of Computer Science} \\
\textit{Georgia Institute of Technology}\\
Atlanta, USA\\
asuprem@gatech.edu}
\and
\IEEEauthorblockN{Rodrigo Alves Lima}
\IEEEauthorblockA{\textit{School of Computer Science} \\
	\textit{Georgia Institute of Technology}\\
	Atlanta, USA\\
	ral@gatech.edu}
\and
%

}

\maketitle

\begin{abstract}

A rapidly evolving situation such as the COVID-19 pandemic is a significant 
challenge for AI/ML models because of its unpredictability. 
The most reliable indicator of the pandemic spreading has been the number 
of test positive cases. 
However, the tests are both incomplete (due to untested asymptomatic cases) and 
late (due the lag from the initial contact event, worsening symptoms, and test results). 
Social media can complement physical test data due to faster 
and higher coverage, but they present a different challenge: 
significant amounts of noise, misinformation and disinformation. 
We believe that social media can become good indicators of pandemic, 
provided two conditions are met. 
The first (True Novelty) is the capture of new, previously unknown, 
information from unpredictably evolving situations. 
The second (Fact vs. Fiction) is the distinction of verifiable facts 
from misinformation and disinformation. 
Social media information that satisfy those two conditions are called live knowledge. 
We apply evidence-based knowledge acquisition (EBKA) approach to collect, 
filter, and update live knowledge through the integration of social media 
sources with authoritative sources. 
Although limited in quantity, the reliable training data from authoritative sources 
enable the filtering of misinformation as well as capturing truly new information. 
We describe the EDNA/LITMUS tools that implement EBKA, integrating social 
media such as Twitter and Facebook with authoritative sources such as WHO and 
CDC, creating and updating live knowledge on the COVID-19 pandemic.
	
\end{abstract}


\section{Introduction}
\label{sec:a}
By definition, epidemics spread rapidly and widely. Furthermore, like many disasters, epidemics also change significantly the environments they invade. As of October 22, 2020, the COVID-19 pandemic has spread to more than 180 countries and regions worldwide, with 42M+ cases and 1.1M+ deaths \cite{who}. COVID-19 has transformed the world into the 'New Normal' with social distancing, travel restrictions, remote work, and online learning. Even with early lockdowns, several countries including the US have recently seen resurgence of cases. 

With several vaccines in phase 3 clinical trials, but none approved for general use, the control of COVID-19 has depended on measures such as shelter-in-place, wearing of masks, and closure of high-risk businesses. These measures are unpopular for human, social, and economic reasons \cite{infodemic}. Entire sectors of economy have suffered, including travel, professional sports, and retail. When applied unevenly, the restraining measures have had mixed results, with a third wave in the USA, both in confirmed cases Figure~\ref{fig:covidcases} and deaths. 

The complexity of COVID-19 pandemic management can be seen in the varied responses in Figure~\ref{fig:covidcases}, with multiple waves at different times for each country. The variability has defied the modeling and control efforts based on the knowledge and assumptions derived from past experiences such as the 2003 SARS outbreak and annual flu seasons. Among other reasons, this complexity has been attributed to the high transmission rate of the SARS-COV-2 virus, and the relatively high number of asymptomatic cases that were contagious, contributing significantly to the spreading of the pandemic. These factors make contact tracing, the main containment tool in past pandemics, ineffective in countries with high infection rates such as the USA, India, and Brazil. 

In an evolving pandemic, timely and reliable information becomes extremely important for decision making at all levels, from the government to the general public. However, there are difficulties in obtaining actual information quickly. First, there are significant technical challenges in data collection, processing, and filtering of truly new information in a timely manner. Second, bad-faith actors generate misinformation and disinformation for monetary and political gains. Third, there are non-technical issues that may obstruct the information flow [infodemic REF]. In this paper, we postpone the non-technical issues and focus only on the two technical challenges.

The first technical challenge, which we call true novelty, is the timely discovery of new information that has never been seen before. This is a difficult challenge for two reasons. First, continuous, real-time monitoring is required for the collection of new information. Second, statistical methods typically classify previously unseen outliers as noise. This happened to unprecedented discoveries such as the Ozone Hole over Antarctica, verified by ground radar data. Retrospective analysis of NASA satellite data shows the phenomenon appeared in satellite data several years before, but the unprecedented (low) values were considered 'physically impossible' by data assimilation algorithms and filtered out \cite{ozone}. These two challenges are handled by the EDNA toolkit and EBKA approach to include corroboration into event discovery (Section~\ref{sec:d}). 

The second technical challenge, which we call fact vs. fiction, consists of the deluge of data in all communications channels on dominant topics such as COVID-19 pandemic, including information, misinformation, and disinformation, a problem often referred to as 'fake news', and 'infodemic' [cite WHO and India paper]. Despite the difficulties of distinguishing the real information from fake news, it is critical for everyone, from policy makers to the general public, to make appropriate decisions based on real information. This is a real threat, since the continued growth of infections and deaths worldwide Figure~\ref{fig:covidcases} suggests that many people may have been influenced by fake news in their decisions and behaviors. Another indication is the recent poll by Pew Research \cite{pew}, which shows only slightly more than half of the US public would be willing to take COVID-19 vaccines. The distinction of facts from non-facts is also handled by the EBKA approach (Section~\ref{sec:d}).

The rest of the paper is organized as follows. Section~\ref{sec:b} outlines the opportunities and potential benefits of collecting high quality social media data towards the tracking of COVID-19 pandemic. Section~\ref{sec:c} outlines the technical challenges as mentioned above, and related work in those areas. Section~\ref{sec:d} describes the technical approach to addressing those challenges, including the EDNA toolkit for real-time social media data collection, and EBKA approach to distinguish facts from fiction. Section~\ref{sec:e} concludes the paper.

\section{Tracking the Pandemic with Live Data}
\label{sec:b}

\subsection{Physical Indicators to Detect and Control the Pandemic}
\label{sec:b1}
\begin{figure*}[t] 
	\centering 
	\includegraphics[width=\linewidth]{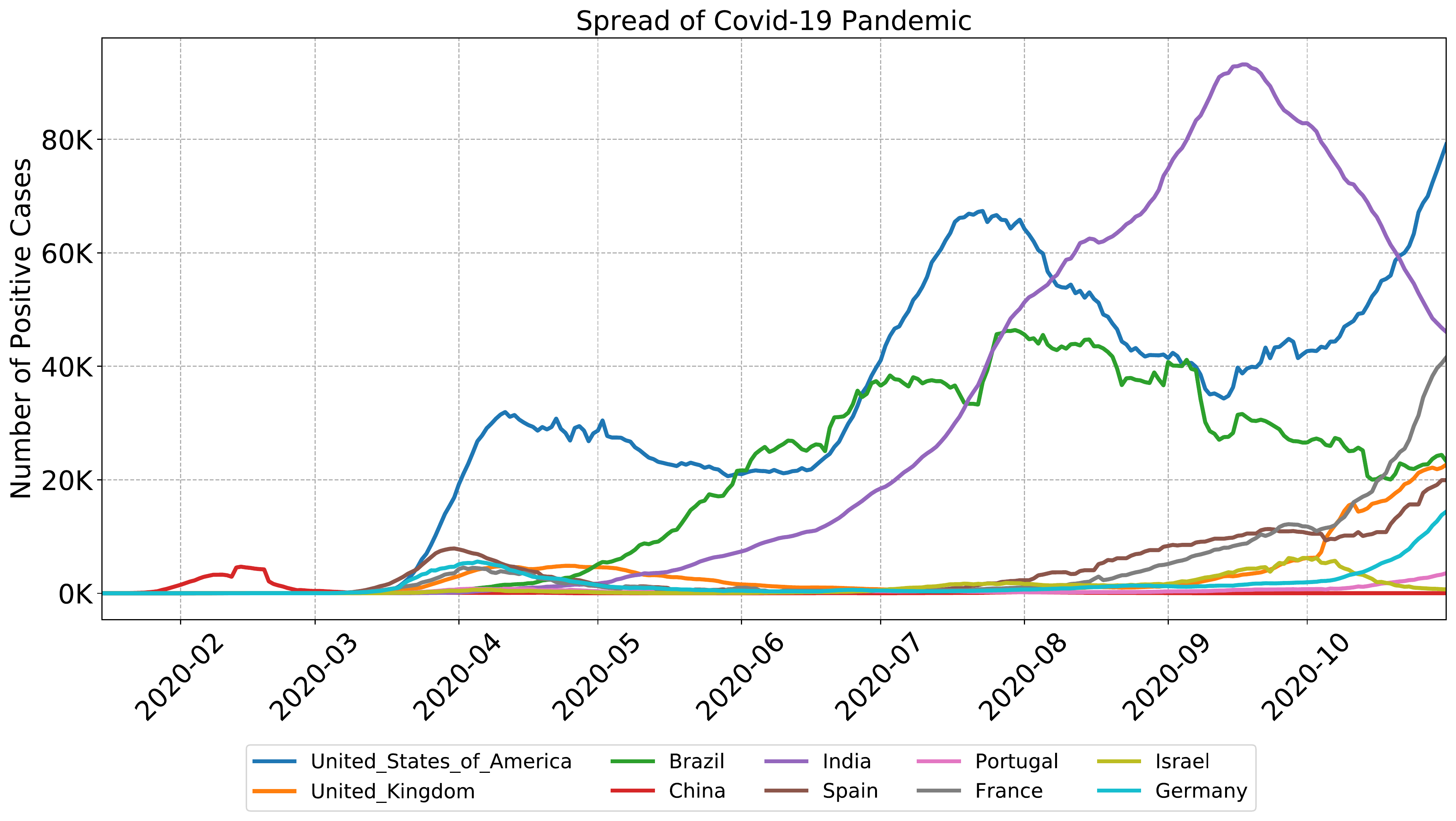}
    \caption{\textbf{Covid Cases}: Test-confirmed new cases in the world (7-day moving average, Oct 31, 2020 \cite{jhu})}
    \label{fig:covidcases}
\end{figure*}

The tracking of COVID-19 pandemic has been accomplished through physical indicators as in past epidemics. The primary indicator is the number of 'test positive' cases, as shown in Figure~\ref{fig:covidcases}. The second indicator is the number of deaths attributed to the pandemic. Together with the number of hospitalizations (often with incomplete reporting), the number of deaths is considered the main indicator of social and economic cost caused by the pandemic, since the asymptomatic and mild cases have less impact on people. The test positive graph is an important predictor since a percentage of positively identified cases will become seriously ill and require hospitalization, and a percentage of these patients will die. 

While very useful as statistical predictor of hospitalization and deaths, unfortunately the test positive cases by themselves have become less effective towards slowing or stopping the spread of the pandemic. The current situation is quite different from previous epidemics, since the test positive cases form the foundation of contact tracing, the main method of effective containment. Contact tracing can be modeled as a graph closure algorithm. Each person is a node, and close contact is an edge connecting the two nodes. By testing all nodes in contact with infected nodes, contact tracing can find all possible candidates of infection and stop future contacts. The key is that the contact tracing effort must overtake the epidemic propagation to stop it successfully. 

In COVID-19 case, due to the high contagion rate $R_0$, a large number of infection cases have quickly overwhelmed the limited number of contact tracers, making it impractical to control the pandemic through traditional contact tracing. Another significant factor is contagion through asymptomatic cases, estimated to be as high as 40\% of total \cite{asymp}, particularly among the young. Strategies that worked for Taiwan and other Asian countries (e.g. aggressive contact tracing through full sharing of patient travel and health histories combined with widespread acceptance of face masks) may not be directly applicable in countries such as the US and European Union for a variety of cultural and political reasons such as privacy concerns. 

To replace manual contact tracing, many mobile apps and platforms have been developed and deployed, including the Google/Apple Privacy-Preserving Contact Tracing API \cite{google,apple}. However, these efforts aimed at automating the contact tracing efforts have been hampered by low rates of adoption. Statistical models estimate the need for an adoption rate of between two-thirds to 90\% of population (depending on the contagion rate $R_0$) for the contact tracing activity to overtake transmission rates successfully. Unfortunately, real world deployments show less than 10\% adoption among mobile phone users \cite{lancet}. The prevailing low adoption rates are due to a variety of reasons, including privacy concerns. As result, the mobile app-based automated contact tracing approach by itself has been insufficient in achieving full control of the pandemic in all deployment efforts.

\subsection{B.2	Tracking Pandemic through Information Propagation}
\label{sec:b2}

In addition to the physical indicators described in Section~\ref{sec:b1}, information through communications media (e.g. social networks) have been explored as alternative or supplementary sources for tracking physical events such as an epidemic. Particularly, social media such as Twitter and WeChat services have proved to be early disseminators of new information, ahead of official reports. In the COVID-19 epidemic, the vast majority of the mild to no symptoms patients have yet to be tested. Consequently, social and online media, including self-reporting, may be the best, and perhaps the only way to gather the missing information on the 80\% of patients.

We acknowledge that some previous attempts such as Google Flu Trends (Section~\ref{sec:c1}) have met serious technical and social challenges such as concept drift (Section~\ref{sec:c2}) and misinformation (Section~\ref{sec:c3}). Our contention is that an information-based pandemic tracking system could provide useful and valuable information if two conditions hold:

\begin{enumerate}[label={\textbf{Condition\arabic*}}, align=left]
\item \lbrack True Novelty\rbrack\, Sufficient quantity of truly new information can be obtained in a timely manner
\item \lbrack Fact vs. Fiction\rbrack\, Factual information of sufficient quality can be separated from misinformation and disinformation
\end{enumerate}

We recognize the above two conditions represent significant challenges vis-\`a-vis current state of the art in ML research. \textbf{Condition 1} (True Novelty) is a challenge that distinguishes our work from classic ML research based on closed data sets such as MNIST and CIFAR. As an illustrative example, models built on pandemic data collected in the US during the first wave (Apr – May of 2020) may not apply directly to the second wave (Aug – Sept), or the incipient third wave (October), due to evolving social, economic, and political environments that necessarily affect communications channels such as social media. While the True Novelty condition may appear obvious in retrospect, it has been obscured by the ML tradition of working (only) within closed data sets collected and annotated in yesteryears.

\textbf{Condition 2} (Fact vs. Fiction) is a challenge recognized by researchers and practitioners in the information security area, ranging from spam detection \cite{spam1,spam2,spam3,spam4} (in various communications channels including email, web, and social media) to fake news \cite{fakenews}. On the COVID-19 pandemic topic, the World Health Organization (WHO) has adopted the term 'infodemic' \cite{infodemic} to describe the rampant spreading of misinformation and disinformation for both monetary and political gains \cite{covidmisinfo}. However, research on, and models developed for historical data sets such as fake news from 2016 US presidential elections \cite{politics}, would not apply to the infodemic problems on the COVID-19 pandemic.

Our technical approach (described in Section~\ref{sec:d}) will address these challenges and it is our contention that the above conditions can be met under appropriate assumptions. Here, we outline a scenario where the propagation of a pandemic such as COVID-19 can be tracked through information propagation under the two conditions. The scenario consists of two matching components: a physical model and an information model. The physical model follows what we have learned about the pandemic:

\begin{enumerate}
\item The most important social, economic, and human cost is due to the high level of hospitalizations and deaths from the pandemic. The deaths data is available, but lags positive tests by up to 30 days. As such, the death statistics have limited predictive power on pandemic propagation. 
\item The most important predictor of deaths consists of the positive data from Figure~\ref{fig:covidcases}. This is particularly the case of people with high risk factors such as age, and chronic diseases such as diabetes. Except for a few countries and regions with community testing, positive tests typically follow the onset of significant symptoms (Up to 10 days after contact event).
\item The traditional method to contain epidemics is contact tracing, where the contact event is reconstructed, and all the persons involved are tested for contagion. A graph closure algorithm catches up with the propagation of virus, enabling the elimination of further contacts and contagion. Contact tracing works when the number of infections is relatively small and infection symptoms are clear. Unfortunately, the COVID-19 pandemic is spreading too fast, and it has up to 40\% of asymptomatic cases that are still contagious. 
\end{enumerate}

The information propagation model of social media follows the physical model closely. Our research is conducted in four stages. In the Stage 1, we collect social media data on the pandemic in general (Section~\ref{sec:d3}) and apply the EBKA approach to filter out misinformation and disinformation. In Stage 2, the resulting data set will be further subdivided into the following groups:

\begin{enumerate}
    \item Deaths and hospitalizations: We will apply unsupervised clustering algorithms using keywords such as \textit{death} and \textit{hospitalization}, as well as supervised ML algorithms to build the first group of social media data. This first group will be correlated with the official death data (group 1.a), and hospitalization data whenever available (group 1.b). 
    \item Positive tests: We will apply again unsupervised and supervised ML algorithms using keywords related to positive tests to build the second group of social media data associated with positive tests. We expect one part of group 2 data to be associated with \textit{hospitalized} (group 2.a) and another part with \textit{non-hospitalized} (group 2.b). 
    \item Symptomatic: We will apply a third time unsupervised and supervised ML algorithms using keywords related to COVID-19 symptoms to build a third group of social media data. We expect one part of group 3 to be associated with positive tests (group 3.a), and another part that concerns only general discussions on symptoms (group 3.b). 
\end{enumerate}

In Stage 3 of our research, we will search for a correlation in space and time between the relevant social media data (groups 1.a and 2.a) with the physical data on deaths, and positive tests. The EDNA toolkit (Section~\ref{sec:d3}) is able to find the location of an event when mentioned in a tweet. The space-time matching of social media data with the physical data of that area is important, since different countries and regions have different time frames in pandemic propagation. The variations among the countries in Figure~\ref{fig:covidcases} (and variations among the states in the US) are indications of the need for localization in the time correlation analysis. 

In Stage 4 of our research, assuming a reasonable correlation between the physical model and information propagation model can be established, we will search for correlation between contact events and discussion of symptoms (group 3) as well as positive tests (group 2). This search will start from known contact events where social distancing was optional, e.g., the Sturgis motorcycle rally and many in-person, crowded election campaign events and rallies. The next step of search will focus on medium scale events that have been reported as super-spreaders, including weddings and church events. The search will continue with friend and family events in the community transmission stage. 
\section{Challenges and Related Work}
\label{sec:c}

\subsection{Associating Physical Events with Social Media}
\label{sec:c1}
There have been several attempts to study the association between physical events with social media and other communications media. Methodologically, many of these studies follow the tradition in big data analytics to post-process the raw data set collected on an event through various data cleaning steps into a closed data set. Then, ML classifiers are generated (either through unsupervised or supervised learning) and tested on the cleaned data set. Due to the heuristics approaches to data cleaning that are often specific to each event, the reported results are often specific to that event. Most importantly, such retrospective data analyses happen after the fact, often years later, making the approach inapplicable to near-real-time responses needed for rapidly changing situations such as the COVID-19 pandemic.

The AI/ML community has adapted to this limitation of retrospective studies by mapping terms and concepts from the real world (e.g., 'real-time' as used by real-time community such as RTSS and RTAS conferences) into the closed context of historical data sets. As a concrete example, the highly cited Sakaki paper \cite{sakaki} (4500+ in Google Scholar as of September 2020) is entitled 'Earthquake shakes twitter users: real-time event detection by social sensors'. However, it is a retrospective study conducted years after the earthquakes. The term 'real-time' in the title refers to the difference between two timestamps: (1) a historical earthquake event, and (2) the moment their model is able to decide on the detection of that earthquake, based on the accumulated tweets from the retrospectively cleaned Twitter log on the earthquake.

An implicit expectation of retrospective studies is that the models trained from past event data would be applicable to similar events in the future. Although this natural expectation remains a valid and important goal for research in this area, the actual ML models developed from past retrospective studies have consistently shown less than robust performance when applied to newer events. This difficulty is not unique to the models developed in \cite{sakaki}. To the best of our knowledge, all retrospectively trained ML models have had limited (and decreasing) accuracy when evaluated on their modeling and prediction of future events. This discussion of challenges in applying ML models trained from retrospectively cleaned data sets (and closed data sets more generally) is due to the heuristic and event-specific rules of data cleaning, the lack of knowledge about true novelty, and concept drift, the topic of next subsection.

\subsection{Concept Drift and Evolution of Reality}
\label{sec:c2}
Big data approaches, including ML models, have been attempted in the prediction of epidemic propagation. An early and well-known example was Google Flu Trends (GFT), which tracked the progression of annual flu epidemic in the USA, using search terms collected from Google search engines. In a 2009 Nature paper \cite{gft}, GFT showed 97\% accuracy in the prediction of flu propagation, using a statistical model derived from 2007-2008 flu season data. However, a gradual shifting of words and language used in online media (including search terms), a phenomenon called concept drift \cite{argo}, caused a steady decay of GFT performance, with prediction errors reaching more than 100\% by 2012 \cite{gftbad}; GFT was officially shut down in 2015.

According to current ML research practices on evaluating ML models within closed data sets such as MNIST\cite{mnist} and CIFAR\cite{cifar}, most of concept drift \cite{survey} papers have restricted their attention to closed data sets in a restricted case called virtual concept drift. A typical technique to handle virtual concept drift is to allocate appropriate weights to members of a teamed classifier, which can adapt to varying subsets of the closed data set. Unfortunately, closed data sets, including the retrospectively cleaned data on events, cannot adapt to environmental changes in the real world, which is happening with the COVID-19 pandemic. Even for the GFT case, from 20/20 hindsight, the growth of GFT error rate was only partially due to changes and evolution of language used to search. Another important factor is that that the reality of flu has also evolved over the years, including new viral treatments such as Tamiflu that became more widely available and affordable. Given non-trivial evolution of reality, virtual concept drift techniques would be unable to maintain accuracy, since the new reality (e.g., the COVID-19 pandemic) did not exist in the original training data.

\subsection{Misinformation and Disinformation}
\label{sec:c3}

Due to the impact of the COVID-19 pandemic, a significant portion of social media has been devoted to the topic. This attention has generated an extraordinary amount of misinformation and disinformation on all topics related to the pandemic, a phenomenon called 'infodemic' \cite{infodemic} by the WHO. Some of the misinformation and disinformation fall into the category colloquially known as 'fake news' \cite{fakenews}, and other items have been classified as pseudo-science. Examples of fake news include rumors about purported cures for the SARS-COV-2 virus (e.g.,chloroquine), and causal agents (e.g., 5G cellphones or Bill Gates caused the pandemic). Examples of pseudo-science include articles written in technical paper style that use genetic sequence analysis to supposedly prove the SARS-COV-2 virus was created in a bio-weapons lab (that belongs to CIA or located in Wuhan, depending on the source).

The presence of misinformation and disinformation in social media has been a well-known problem, as old as social media themselves, and preceded by similar problems in other communications media such as email and web spams. However, the scale, persistence, and sophisticated of COVID-19 infodemic is unmatched, due partially to the scale and persistence of the pandemic itself. As an illustrative example of the challenge, consider the application of ML techniques in the sentiment analysis area \cite{sentiment}. Based on statistical clustering, sentiment analysis is good at finding strong opinions, but it would have difficulties distinguishing actual facts from inaccurate or malicious (strong) opinions. This problem has been exacerbated by fake accounts that create overwhelming quantity of fake news through repetition (e.g., similar Facebook postings or retweeting).

Due to the difficulties in the interpretation of opinions and their ease of change, we restrict our attention to verifiable facts, which form the core part of the physical disaster management area, including pandemics such as COVID-19. Unlike opinions, the veracity of facts is unaffected by how loud the shouting is. Instead, facts are published by reputable or authoritative sources, and corroborated through a continuous verification process by independent fact checkers. In our research, we will rely on authoritative sources such as CDC \cite{cdc} and WHO situation reports\cite{who} on pandemic data including deaths and positive tests. In addition, we also use reputable sources (that employ corroboration before publication) such as the Johns Hopkins information center \cite{jhu} as well as reputable news sources including NY Times \cite{nyt} and CNN \cite{cnn}.
\section{Technical Approach}
\label{sec:d}

\subsection{True Novelty on Pandemic in Social Media}
\label{sec:d1}

Towards the satisfaction of Condition 1 on True Novelty, we have developed a set of tools \cite{multilingual} for data collection, classifier training, and data analytics for the near-real-time detection of landslides in the LITMUS project \cite{litmus}. The tools are highly customizable for a variety of topics, e.g., the LITMUS tools have been successfully used to collect and analyze data on wildfires. The tools include data collectors from Twitter and Facebook as primary sources, and reputable news sources such as New York Times and CNN.com as corroborative sources. The LITMUS tools have evolved through several iterations, include the ASSED \cite{assed} and EDNA toolkits \cite{edna}, which achieve faster deployment, fault tolerance, and end-to-end management. 

EDNA (and LITMUS) tools collect streaming social media data from channel-specific APIs, e.g. the Twitter Streaming API. As an example, the Twitter API sends tweets that satisfy selection criteria defined by topic keywords. Similarly, newspaper APIs send news articles matching the selection criteria. For the data collection on the COVID-19 pandemic, topic keywords include: \textit{coronavirus}, \textit{covid-19}, \textit{ncov-19}, and \textit{pandemic}, among others. The near-real-time collection of data from live social media sources is the first step of live knowledge aggregation process.

The second step concerns the issues of concept drift \cite{survey} and evolution of reality discussed in Section~\ref{sec:c2}. The discussions about the pandemic can shift abruptly as new topics are introduced, e.g., when new drugs and vaccines are announced, and/or systematic misinformation and disinformation campaigns are initiated. Although new keywords can be added manually, the crucial period of true novelty detection (at the beginning when the novelty was introduced) may have passed and the new information lost. EDNA adapts to concept drift in two ways: (1) by automating the augmentation of topic keywords, and (2) by leveraging social network tracking of popular postings.

First, as live social media data items are collected from existing keywords, EDNA applies clustering algorithms to search for new popular keywords. Our assumption is that both positive new topics (e.g., new treatments) and new campaigns of fake news associated with the pandemic would still contain some of the pandemic-related keywords, at least at the beginning, due to the requirements of current search algorithms. This is because under true novelty, this correlation might disappear as the sources of these posts change communities. Capturing nascent keywords allows us to follow the evolution of new topics. For example, keywords such as: \textit{mask}, \textit{bioweapon}, and \textit{bill gates} were highly correlated in the early stages of the pandemic. After a few weeks or months, the correlation decreased significantly since the new keywords have acquired their own social context. Capturing the keywords in the early stages allowed us to continue collecting data through concept drift, e.g., new tweets containing only the keyword mask and omitting the original keywords \textit{coronavirus}, \textit{covid-19}, \textit{ncov-19}, or \textit{pandemic}.

Second, EDNA also benefits from the popular or viral tweets detected by social networks (using their own algorithms and human moderators). Many of trending tweets would contain the same topic keywords that EDNA is tracking. However, other trending tweets can contain images and memes, or neologisms that relate to the pandemic, but having low textual correlation with current keywords. Example of relevant neologisms include 'infodemic' (a legitimate concept to describe the misinformation campaigns on the pandemic, being popularized by WHO), and \textit{plandemic} (a fake news campaign that emerged in June 2020), used to denote a particular anti-vaccine conspiracy theory.

\subsection{True Novelty on Pandemic in Social Media}
\label{sec:d2}
The discussion in Section~\ref{sec:d1} observes the appearance of both legitimate new information and misinformation when tracking new data items that belong to true novelty category. To overcome the limitations of traditional ML classifiers trained from closed data sets, we propose the application of Evidence-Based Knowledge Acquisition (EBKA) approach \cite{ebka} to integrate noisy social media data such as Twitter, Facebook, and Weibo with authoritative sources such as WHO and CDC reports \cite{who,cdc} to distinguish verifiable facts from fake news. Initially applied in the LITMUS landslide information service \cite{litmus}, EBKA has demonstrated successful 

The EBKA approach was initially developed in the LITMUS project to filter the overwhelming amount of noise in social media on landslide disasters. The main challenge is that only about 5-10\% of tweets that contain the keyword 'landslide' actually referred to landslide disasters, with the majority of references on results of elections as well as soccer matches. EBKA distinguishes true novelty in social media from misinformation through an automated integration of primary sources (social media such as Twitter and Facebook) with authoritative sources (reputable news sources such as NY Times and CNN). 

Since the authoritative sources only report on large disasters with news reports and many tweets, it is uncertain whether the classifiers trained from large-disaster data would be able to detect small-scale landslides that have few tweets. Fortunately, their complementary nature enabled a good combination: primary sources having high coverage with high noise levels, and authoritative sources having high reliability and low coverage levels. LITMUS results show that the high reliability of ground truth from authoritative sources on a few large disasters (between 1-5\% of detectable landslides) combines well with the relatively good reporting on each large landslide. The result consists of high quality teamed deep learning (DL) classifiers that become capable of detecting the smaller new landslides that have lower signal-to-noise ratio. These EBKA teamed classifiers become capable of recognizing a total of more than 10 times real landslide disasters with high accuracy.

The automated EBKA process to generate high quality teamed classifiers from true novelty helps satisfy Condition 2 on the distinction of verifiable facts. By integrating authoritative sources with primary sources, EBKA becomes capable of recognizing true novelty information on the pandemic;  the EDNA tools leverage authoritative sources such as WHO, CDC, and JHU COVID Information Center to capture verifiable facts on the pandemic. At an abstract level, the continuously generated true novelty information on verifiable facts is called \textit{live knowledge} \cite{ebka}, which contributes to the opportunities outlined in Section~\ref{sec:b}. Specifically, the continuous capture of live knowledge on the COVID-19 pandemic can improve substantially the timely study of its spread and effective countermeasures.

Like LITMUS, EDNA will use teamed classifiers to identify physical event clusters from social media. For example, multiple tweets that refer to an event (e.g., an outbreak of positive tests) at the same time and location can be grouped into a tentative event cluster. Event detection from social media is fraught with noise: location extraction is somewhat coarse, and this limits specific event detection. To filter out such noise and collect information with verifiable facts, LITMUS integrates authoritative sources such as news articles or other reputable sources to (a) retroactively correct any inaccuracies in event clustering, and (b) update event cluster detection models, e.g., whether a cluster is a true novel event, or fake news. 

Retroactive correction involves progressively refining the previous decision on older event clusters from social media that have received more authoritative information, e.g., from fact finding websites. In case of COVID-19, we ensure each event cluster found in social media (e.g. tweets mentioning the Sturgis motorcycle rally with coronavirus case related keywords such as 'crowd' or 'gathering') has supporting evidence from authoritative sources (e.g. news articles referring to the Sturgis rally as a super-spreading event at a later date).

EDNA also performs methodical rolling model updates to refine event cluster detection, particularly in the important decisions such as whether an event is real or fake news. Event clusters are detected with keyword filters plus topic modeling that extracts trending topics related to event keywords. Refining event cluster detection requires identifying new keywords that may be related to the pandemic, such as 'bleach' after the President's press conference on April 24, and updating the topic modeler to include these new keywords. Once a new coronavirus related trend is detected, EDNA retrains the event cluster detector with existing event cluster tweets and the new trend tweets to add them to the teamed classifier.

The rolling model updates provides additional evidence from authoritative sources to substantiate a true novelty cluster as real event. In contrast, if negative evidence is found (e.g., fact finding websites indicating a topic to be fake news), EDNA  reduces the weight of sub-models related to those event cluster tweets in the teamed classifier. This progressive substantiation process gave name to EBKA (evidence-based knowledge acquisition): the teamed classifiers gains more information on true novelty through the accumulation of supporting or contradicting evidence from authoritative sources. Since there should be no controversies on verifiable facts, we expect the process to converge quickly and the occurrence of fact settled.

Applying EBKA approach to the COVID-19 pandemic, EDNA has collected about 600GB of social media data from several social networks as outlined in the Section~\ref{sec:d3}, with over 600M tweets.

\subsection{Continuously Collected Primary Dataset (Condition 1)}
\label{sec:d3}

We have collected the EDNA-Covid dataset since January 25, 2020 using Twitter’s streaming API. Over time, we have also enriched our dataset with other similar datasets, such as \cite{chen} and \cite{coviddata}. We use the Twitter Sampled Stream API. This API provides a real-time stream of 1\% of all tweets. In EDNA, we collect this stream with a highly-available cluster of ingest processes. During our data ingest, we perform our keyword extraction to identify coronavirus related tweets, with the following keywords: \textit{corona}, \textit{covid-19}, \textit{ncov-19}, \textit{pandemic}, \textit{mask}, \textit{wuhan}, and \textit{virus}. To capture Chinese social data, we also include these keywords in Mandarin. We initially included the keyword \textit{china} during data collection in January and February, but decided to omit the phrase since it introduced significant noise, and any tweets with the keyword that were relevant to coronavirus already include the above keywords.

Even with 1\% of the Twitter stream, we are able to collect a large scale dataset of tweets. We show in Table 1 the tweets collected since January. We converted to a highly-available cluster of ingest processes near the end of June to improve our data collection and reduce instances of dropped tweets. We also updated our keyword filtering approach to keep tweets that are retweets of matching tweets.

\begin{table}[h]
	\centering
    \caption{Per-month tweet counts for EDNA-Covid}
    \label{tab:counts}
    \begin{tabular}{lr}
    \hline
    \textbf{Month} & \textbf{No. Tweets} \\ \hline
    2020-01        & 8,714,684           \\
    2020-02        & 25,553,003          \\
    2020-03        & 31,564,785          \\
    2020-04        & 25,498,020          \\
    2020-05        & 26,895,960          \\
    2020-06        & 99,415,221          \\
    2020-07        & 112,215,578         \\
    2020-08        & 113,543,567         \\
    2020-09        & 103,454,256         \\ \hline
\end{tabular}
\end{table}

Our data is skewed towards English language tweets, as we show in Table 1 with the top 5 language categories. We also included Chinese and Japanese tweets with keywords in the corresponding languages; including Chinese keywords nets us ~25K tweets per month, which is less than 0.1\% of the collected tweets, and including Japanese keywords adds ~50K tweets per month. This includes enrichment with tweets from \cite{chen,coviddata}.

\begin{table}[h]
	\centering
    \caption{Top 5 languages in EDNA-Covid dataset}
    \label{tab:language}
    \begin{tabular}{lrr}
    \hline
    \textbf{Language}                                                                 & \textbf{No. Tweets} & \textbf{Pct Total} \\ \hline
    English                                                                           & 395,109,343         & 63.4\%             \\
    Spanish                                                                           & 76,653,705          & 12.3\%             \\
    \begin{tabular}[c]{@{}l@{}}South Asian \\ (Indonesian, Javan, Malay)\end{tabular} & 23,681,633          & 3.8\%              \\
    French                                                                            & 21,812,030          & 3.5\%              \\
    Portuguese                                                                        & 19,942,427          & 3.2\%              \\ \hline
    \end{tabular}
\end{table}

As a starting point for primary source data collection, we have created new Twitter queries with keywords such as \textit{coronavirus}, \textit{covid19}, \textit{novel coronavirus}, \textit{outbreak}, \textit{quarantine}, \textit{sars-cov-2}, \textit{hubei}, and \textit{wuhan}; these keywords include both static and dynamically updated keywords as described in Sections~\ref{sec:d1} and ~\ref{sec:d2}. Three sample tweets are included in the following table:

\begin{table}[h]
	\centering
	\caption{Sample tweets collected with EDNA}
	\label{tab:tweets}
	\begin{tabular}{l}
	\hline
	\begin{tabular}[c]{@{}l@{}}\{ 	"created\_at": "Sat Feb 29 18:59:56 +0000 2020", \\
									"id": 1233829273691049984, "text": "Coronavirus \\
									will spread in California, health officials say: \\
									'It's already out of the bag' https:\textbackslash{}\\
									/\textbackslash{}/t.co\textbackslash{}/YHBt1myH5X \\
									\#uncategorized \#feedly … \}\end{tabular}
									\\ \hline
	\begin{tabular}[c]{@{}l@{}}\{ 	"created\_at": "Sat Feb 29 18:59:57 +0000 2020", \\
									"id": 1233829277067595779, "text": "President \\
									Trump on \#coronavirus: \textbackslash{}u201cIts a \\
									tough one but a lot of progress has been made\\
									\textbackslash{}u201d - 22 cases in US but one die\\
									\textbackslash{}u2026 https:\textbackslash{}/\\
									\textbackslash{}/t.co\textbackslash{}/Siw30dZqQY“ … \}\end{tabular} \\ \hline
	\begin{tabular}[c]{@{}l@{}}\{ 	"created\_at": "Sat Feb 29 18:59:57 +0000 2020", \\
									"id": 1233829277914877953, "text": "Window of \\
									opportunity for containing coronavirus rapidly \\
									closing. https:\textbackslash{}/\textbackslash{}\\
									/t.co\textbackslash{}/WZ7joAxzqt“ ..\}\end{tabular}
									\\ \hline                                                                          
	\end{tabular}
	\end{table}

	This sustained increase in online engagement in reference to a single event provides an unprecedented insight into a slew of areas in natural language processing, such as social communication modeling, credibility analysis, topic modeling, and fake news detection. Our EDNA-Covid dataset, which contains over 600M tweets from over 10 languages, would be an excellent source for research into the social and language dynamics of the pandemic. Our dataset demonstrates concept drift, making it ideal for testing streaming models of analytics.

	Data exhibits concept drift when its underlying distribution changes over time, usually over several years. Under concept drift, machine learning models and conventional offline analytics will degrade as their prediction data desynchronized from their training data model. Concept drift is a natural part of real data; several examples of drift abound in nature, from changing seasons, which can degrade performance of computer vision systems, to lexical drift \cite{lexical}, which can degrade performance of NLP models over different geographical regions. An important requirement inn concept drift research is data that exhibits such drift to enable development and testing of drift detection and adaptation mechanisms.

	With EDNA-Covid, we present a dataset that exhibits concept drift. The online discourse on the Covid-19 pandemic has taken root in a dizzying array of online communities, such as sports \cite{sports}, academia \cite{academic}, and politics \cite{politics}. This allows us a firsthand look at a real-world example of concept drift as the online conversations change over time to accommodate new actors, knowledge, and communities. This yields a high-volume, high-velocity data stream with noise and drift as the underlying conversations about the pandemic transition from confusion to information to misinformation \cite{misinformation} and today, with the US election nearing, disinformation \cite{disinformation}.

\begin{figure}[t] 
		\centering 
		\includegraphics[width=\linewidth]{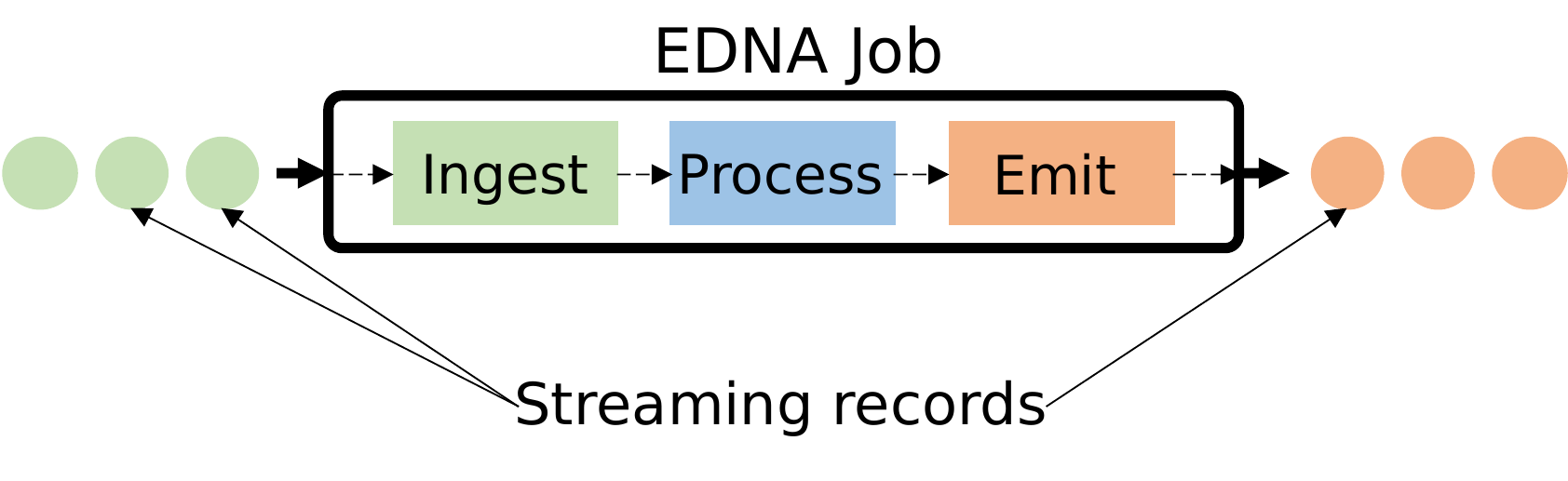}
		\caption{\textbf{An EDNA Job}}
		\label{fig:ednajob}
	\end{figure}

	\paragraph{EDNA} We will briefly describe our EDNA toolkit here. 
	\edna is an end-to-end streaming toolkit for ingesting, 
	processing, and emitting streaming data. 
	\edna{'s} initial use was a test-bed for studying concept drift 
	detection and recovery. 
	Over time, it has grown to a toolkit for stream analytics. 
	We are continuing to work on it to mature it for production clusters. 
	The central abstraction in \edna is the \textit{ingest-process-emit} loop, 
	implemented in an \ednajob. 
	We show an \ednajob in Figure~\ref{fig:ednajob}. 
	Each component of the loop in an \ednajob is an abstract primitive in \edna 
	that is extended to create powerful operators. 
	\begin{itemize}
	\item \textbf{Ingest primitives} consume streaming records. 
	\item \textbf{Process primitives} implement common streaming transformations 
	such as map and filter \cite{flink}. Multiple process primitives can be chained in the same job. 
	\item \textbf{Emit primitives} generate an output stream that can be sent to a storage sink, 
	such as a SQL table, or to another \ednajob.
	\end{itemize}
	
	The \edna stack consists of four layers: deployment, runtime, APIs, and plugins.
	\edna can be deployed on a local machine for single jobs or on clusters managed by orchestrators 
like Kubernetes for multiple jobs in a streaming application. 
On a cluster deployment, \edna uses Apache Kafka \cite{kafka}, a durable message broker 
with built-in stream playback to connect jobs, and Redis \cite{redis} to share information 
between jobs. 
The \edna runtime manages and executes jobs on the applied deployment. 
\ednajob{s} use the ingest, process, and emit APIs to implement the 
\textit{ingest-process-emit} loop, with the appropriate plugin for complete the job.

	\paragraph{Dataset Release.} 
	Due to Twitter TOS regarding release of tweets, we are releasing the dataset to the public through a registration method. We have provided a form at \url{https://forms.gle/dFYhuMzyPMunY17H9} for dataset requests. We have released an alpha version of EDNA at \url{https://github.com/asuprem/edna} and a sample of the dataset at \url{https://github.com/asuprem/
	EDNA-Covid-Tweets}. We deploy long-running stream processing applications with EDNA on Kubernetes. In this case, we deploy LITMUS tools for data collection and classification on EDNA.

\subsection{Factual Dataset with True Novelty (Condition 2)}
\label{sec:d4}

We now describe our steps to address Condition 2 for extraction of the factual dataset from the raw EDNA-Covid dataset. The factual dataset consists of social media post clusters that can indicate changes in the pandemic's spread. We extract the factual dataset with an EDNA application. The EDNA application identifies tentative location for the tweet, possible misinformation or disinformation within the text, and any ties to credible and authoritative sources, e.g. CDC, WHO, JHU, New York Times, or other news organizations:

\begin{itemize}
	\item \textbf{Location Extraction:} Identifying location from tweets or social media is difficult since twewets contain 'short-text', which lacks context for most NLP tools. We accomplish this by using off-the-shelf NLP tools like Stanford NER for the easy cases. Simultaneously, we record any detected locations in a short-term cache. For any short-text where off-the-shelf NER cannot fild locations, we use our short-term cache as a substring match against the text to identify any locations. Under EBKA, we also integrate knowledge from authoritative sources. For any new cases detected reported by CDC and WHO, we add locations for those cases to our short-term cache as well.
	\item \textbf{Misinformation Extraction}: We obtain a collection of	misinformation keywords from Wikipedia \cite{wikipedia} and from \cite{covidmisinfo}. We then use these keywords to filter tweets that contain these keywords, which include \textit{bioweapon} and \textit{plandemic}. We continuously update our misinformation keywords from these sources. To detect new misinformation keywords that may not exist on \cite{wikipedia} or \cite{covidmisinfo}, we track tentative keywords associated with existing misinformation that may be trending. This is because new types of misinformation may try to 'piggyback' on existing trends to quickly gain an audience. 
	\item \textbf{Authoritative Source}: We track a list of authoritative sources and their posts. Any media from these sources, e.g. \cite{cdc,who,nyt,cnn,jhu} is automatically counted as factual information.
\end{itemize}

\begin{figure*}[t] 
	\centering 
	\includegraphics[width=\linewidth]{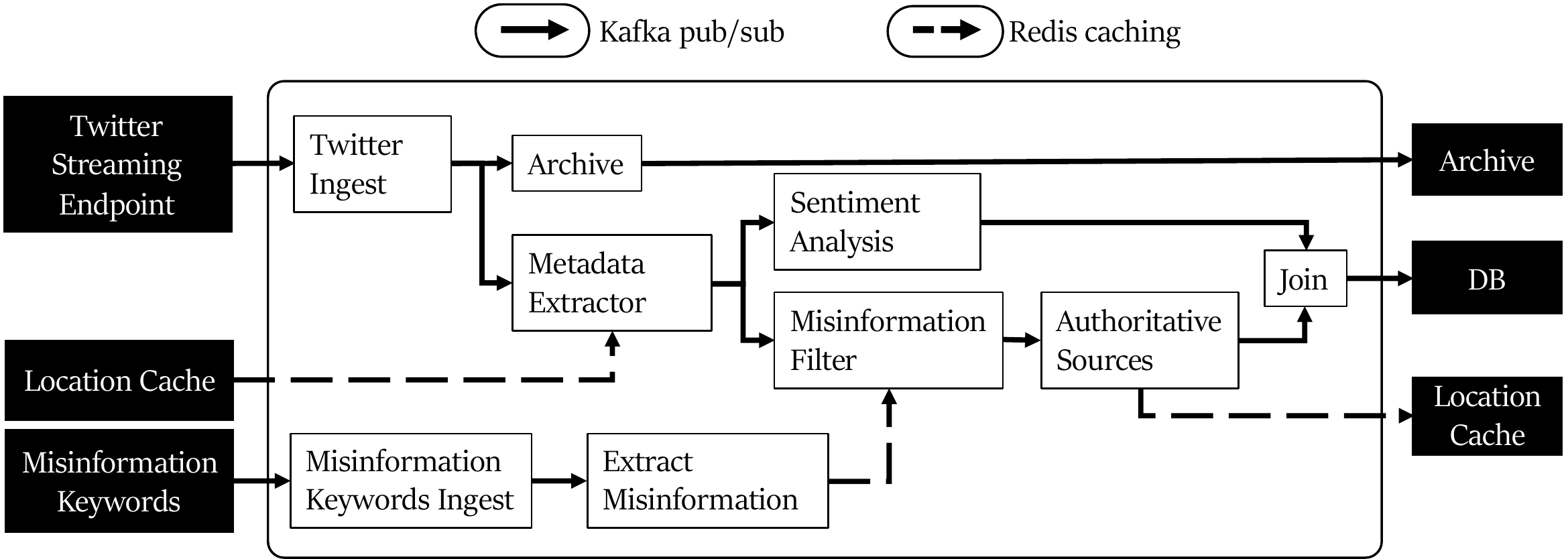}
    \caption{\textbf{EDNA Application}: Our EDNA application to address Condition 2 to extract factual data from the stream.}
    \label{fig:ednaapp}
\end{figure*}

Our EDNA application is shown in Figure~\ref{fig:ednaapp}, where each process within the application is an EDNA Job. We describe key jobs here:

\begin{enumerate}
	\item \textbf{Twitter Ingest}: This job connects to the Twitter v2 sampled stream endpoint, available at \cite{tweets}. 
This API provides a real-time stream of 1\% of all tweets. Each raw object is archived to disk. We have described this dataset in Section~\ref{sec:d3}.
\item \textbf{Metadata extractor}: This job extracts the tweet object from the streaming record and performs some 
data cleaning in discarding malformed, empty, or irrelevant tweets. Tweets without the relevant coronavirus keywords are tagged as possibly irrelevant. Keywords include: 
\textit{coronavirus}, \textit{covid-19}, \textit{ncov-19}, \textit{pandemic}, \textit{mask}, \textit{wuhan}, and \textit{virus}.  This job also performs location extraction using a combination of off-the-shelf Stanford NER and our short-term location cache. 
\item \textbf{Sentiment analysis}: We use an off-the-shelf tweet sentiment analysis model from 
\cite{sentiment} to record text sentiment. 
We plan to replace this with an EDNA application that will automatically generate and 
retrain a sentiment analysis model with data from Twitter's own streaming sentiment operators.
\item \textbf{Misinformation Keywords Ingest}: We obtain a collection of misinformation keywords from Wikipedia \cite{wikipedia} and from \cite{covidmisinfo} as described earlier. This job regularly queries sources for new keywords.
\item \textbf{Extract Misinformation}: This job parses the misinformation sources from \textbf{Misinformation Keyword Ingest}. For example, it extracts keywords from headlines in the \textit{Conspiracy} section for \cite{wikipedia}. All keywords are updated in an internal cache for the \textbf{Misinformation Filter} job.
\item \textbf{Misinformation Filter}: We group 1 minute's worth of tweets for faster misinformation 
keyword checking and to record misinformation keyword statistics on a per-minute window. This job checks whether the grouped tweet objects contain any 
of the misinformation keywords extracted by the \textbf{Extract Misinformation} job and regularly updates its own cache of keywords. Tweets are tagged if they contain misinformatioon keywords.
\item \textbf{Authoritative Sources}: Tweets and content from authoritative sources are tagged with his job. We keep track of authoritative sources as described, including CDC, WHO, NYT, etc.
\end{enumerate}

\section{Vision}
\label{sec:e}

A rapidly evolving situation such as the COVID-19 pandemic is a significant challenge for human decision makers and AI/ML models because of its unpredictability. The most reliable indicator of the pandemic spreading has been the number of test positive cases, but those indicators suffer from being “too few, too late”. The tests are incomplete, since asymptomatic cases (estimated at up to 40\% of total) usually remain untested, and they lag the initial contact by several days, since the symptoms arise a few days later, and the test results often take another couple of days. Additional indicators and predictors of pandemic spread can have a significant impact. 

Social media can complement physical test data due to the faster and higher coverage of social media. However, social media also contain significant amounts of noise, misinformation and disinformation, making them less reliable. In addition, technical issues such as concept drift have rendered ML techniques less effective in rapidly evolving situations. Our hypothesis is that social media can become good indicators and perhaps predictors of pandemic, provided two conditions are met. The first (True Novelty) is the capture of new, previously unknown, information from unpredictably evolving situations. The second (Fact vs. Fiction) is the distinction of verifiable facts from misinformation and disinformation. Social media information that satisfy those two conditions are called live knowledge. 

We apply evidence-based knowledge acquisition (EBKA) approach to collect, filter, and update live knowledge on the spread of COVID-19 epidemic. EBKA integrates primary social media sources such as Twitter and Facebook with authoritative sources such as WHO and CDC. Although the authoritative sources have limited coverage, EBKA is able to use them to generate highly reliable training data and extensible teamed classifiers capable of filtering out misinformation (Condition 2) as well as capturing truly new information (Condition 1). EBKA has been demonstrated to be effective in the LITMUS landslide information system, and we are applying EBKA in the tracking of COVID-19 pandemic information with promising results. As the US recorded 88K new cases in a single day (October 29, 2020), we can use all the help we can get.

\section*{Acknowledgment}
This research has been partially funded by National Science Foundation by CISE/CNS (1550379, 2026945, 2039653), SaTC (1564097), SBE/HNDS (2024320) programs, and gifts, grants, or contracts from Fujitsu, HP, and Georgia Tech Foundation through the John P. Imlay, Jr. Chair endowment. Any opinions, findings, and conclusions or recommendations expressed in this material are those of the author(s) and do not necessarily reflect the views of the National Science Foundation or other funding agencies and companies mentioned above.

\bibliographystyle{IEEEtran}
\bibliography{main}

\end{document}